\documentclass[runningheads]{cl2emult}
\usepackage[dvips]{epsfig}

\usepackage{makeidx}  
\usepackage{graphicx} 
\usepackage{subeqnar} 
\usepackage{multicol} 
\usepackage{physbb}   
\makeindex            

\newcommand{\simgt}{\,\rlap{\lower 3.5 pt \hbox{$\mathchar \sim$}} \raise 1pt
 \hbox {$>$}\,}
\newcommand{\simlt}{\,\rlap{\lower 3.5 pt \hbox{$\mathchar \sim$}} \raise 1pt
 \hbox {$<$}\,}

\def\Journal#1#2#3#4{{#1} {\bf #2} (#4) #3}
\def\NPB{{Nucl.~Phys.} {\bf B}}
\def\PLB{{Phys.~Lett.} {\bf B}}
\def\PRL{Phys.~Rev. Lett.}
\def\PRD{{Phys.~Rev.} {\bf D}}
\def\ZPC{{Z.~Phys.} {\bf C}}
\def\EPJC{{Eur.~Phys.~J.} {\bf C}}

\begin{document}

\thispagestyle{empty}

\begin{center}
\centerline{\hfill  MPI/PhT/99--042}
\centerline{\hfill  September 1999}
\vskip1.5cm
\renewcommand{\thefootnote}{\fnsymbol{footnote}}
{\huge Forward Jet Production at HERA\footnote[5]{Invited talk at the
Ringberg Workshop {\it New Trends in HERA Physics 1999}, Ringberg
Castle, Tegernsee, Germany, 20 May--4 June 1999 }} 
\renewcommand{\thefootnote}{\arabic{footnote}}
\vskip1.cm
{\Large B.~P\"otter \\
Max-Planck-Institut f\"ur Physik (Werner-Heisenberg-Institut),\\
F\"ohringer Ring 6, 80805 Munich, Germany \vspace{2mm} \\
e-mail: poetter@mppmu.mpg.de }

\vspace{2cm}

{\Large Abstract} 

\vspace{.7cm}

\begin{minipage}[t]{11cm} 
We discuss forward jet production data recently published by the H1
and ZEUS collaborations at HERA. We review how several Monte-Carlo models 
compare to the data. QCD calculations based on the BFKL formalism and
on fixed NLO perturbation theory with and without resolved virtual
photons are described.
\end{minipage}
\end{center}
\cleardoublepage
\setcounter{page}{1}

\title*{Forward jet production at HERA}
\author{Bj\"orn P\"otter}
\institute{Max-Planck-Institut f\"ur Physik, F\"ohringer 
  Ring 6, 80805 M\"unchen, Germany}
\maketitle

\begin{abstract}
We discuss forward jet production data recently published by the H1
and ZEUS collaborations at HERA. We review how several Monte-Carlo models
compare to the data. QCD calculations based on the BFKL formalism and
on fixed NLO perturbation theory with and without resolved virtual
photons are described.
\end{abstract}

\section{Introduction}

It is by now firmly established that the proton structure function
$F_2(x,Q^2)$ shows a steep rise in the small $x$-Bjorken region, i.e.,
below $x=10^{-3}$ \cite{sxH1,sxZeus}. This rise is compatible with
DGLAP evolution \cite{dglap}, where $F_2$ is fitted at a fixed input
scale $Q_0^2$ and then evolved up to $Q^2$ by summing up $\ln Q^2$
terms. In the small $x$ region an alternative way to describe the
evolution might be to sum up $\ln (1/x)$ terms, as it is done in the
BFKL approach \cite{BFKL}. The BFKL equation actually directly
predicts the behaviour of $F_2$ as a function of $x$. An attempt to
unify these two approaches is the CCFM evolution equation \cite{ccfm},
which reproduces both, the DGLAP and the BFKL behaviour, in their
respective regimes of validity. 

\begin{figure*}[ttt]
 \begin{picture}(120,180)
  \put(110,5){\psfig{file=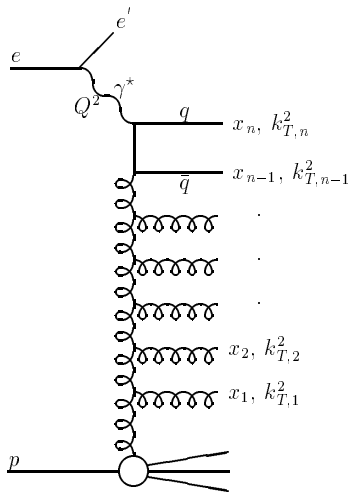,width=4.5cm}} 
 \end{picture}
 \caption{Evolution of the transverse momenta $k_{T,i}$ and the
  longitudinal momentum fractions $x_i$ along the ladder contributing
  to DIS jet production.}
\end{figure*}

The existing data on $F_2$ do not allow to unambigously determine
whether the BFKL mechanism is needed in the $x$-range covered by
HERA. Therefore, alternatively the cross section for forward jet
production in deep inelastic scattering (DIS) has been proposed as a
particularly sensitive means to investigate the parton dynamics at
small $x$ \cite{x1}. The proposal is based on the observation that the
DGLAP and BFKL equations predict different ordering of the transverse
momenta $k_{T,i}$ along the parton cascade developing in DIS jet
production, see Fig.~1. While the DGLAP equation predicts a strong
$k_{T,i}$ ordering and only a weak ordering in the longitudinal
momentum fractions $x_i$, i.e.,
\begin{equation}
 Q^2 = k^2_{T,n} \gg \ldots \gg k^2_{T,1} \qquad \mbox{and} \qquad
 x = x_n < \ldots < x_1
\end{equation}
the BFKL approach predicts a strong ordering in the $x_i$, but no
ordering in $k_{T,i}$. The idea is now to observe jets at very small
$x$, with $x\ll x_1$ in the forward region with $Q^2\simeq
k_{T,1}^2$. In this way the DGLAP evolution is suppressed, whereas the
BFKL evolution is left active. 

The forward jet cross section has been measured recently at HERA
\cite{x5,x6} and in the following I will review the various
attempts that have been made to describe this data.

\section{Data and Monte-Carlo models}

The H1 and ZEUS collaborations have measured forward jet cross
sections at small $x$ for rather similar kinematical
conditions \cite{x5,x6}. The jet selection criteria and kinematical
cuts are summarized in Tab.~1. 

\begin{table}[bbb]
\caption{Forward jet selection criteria by H1 and ZEUS}
\begin{center}
\renewcommand{\arraystretch}{1.4}
\setlength\tabcolsep{5pt}
\begin{tabular}{cc}
\hline\hline
\makebox[4.cm][c]{H1 cuts} &  \makebox[4.cm][c]{ZEUS cuts}\\ \hline 
$E_{e}^{\prime} > 11$ GeV & $E_{e}^{\prime} > 10$ GeV \\
$y_e > 0.1$ & $y_e > 0.1$ \\
$E_{T,jet} > 3.5$ ($5$) GeV & $E_{T,jet} > 5$ GeV \\
$1.7 < \eta_{jet} < 2.8$ & $\eta_{jet} < 2.6$ \\ 
$0.5 < E_{T,jet}^{2} / Q^{2} < 2$ & $0.5 < E_{T,jet}^{2} / Q^{2} < 2$ \\ 
$x_{jet} > 0.035$ & $x_{jet} > 0.036$ \\ \hline \hline
\end{tabular}
\end{center}
\label{Tab1a}
\end{table}

\begin{figure*}
 \begin{picture}(120,160)
  \put(-9,-25){\psfig{file=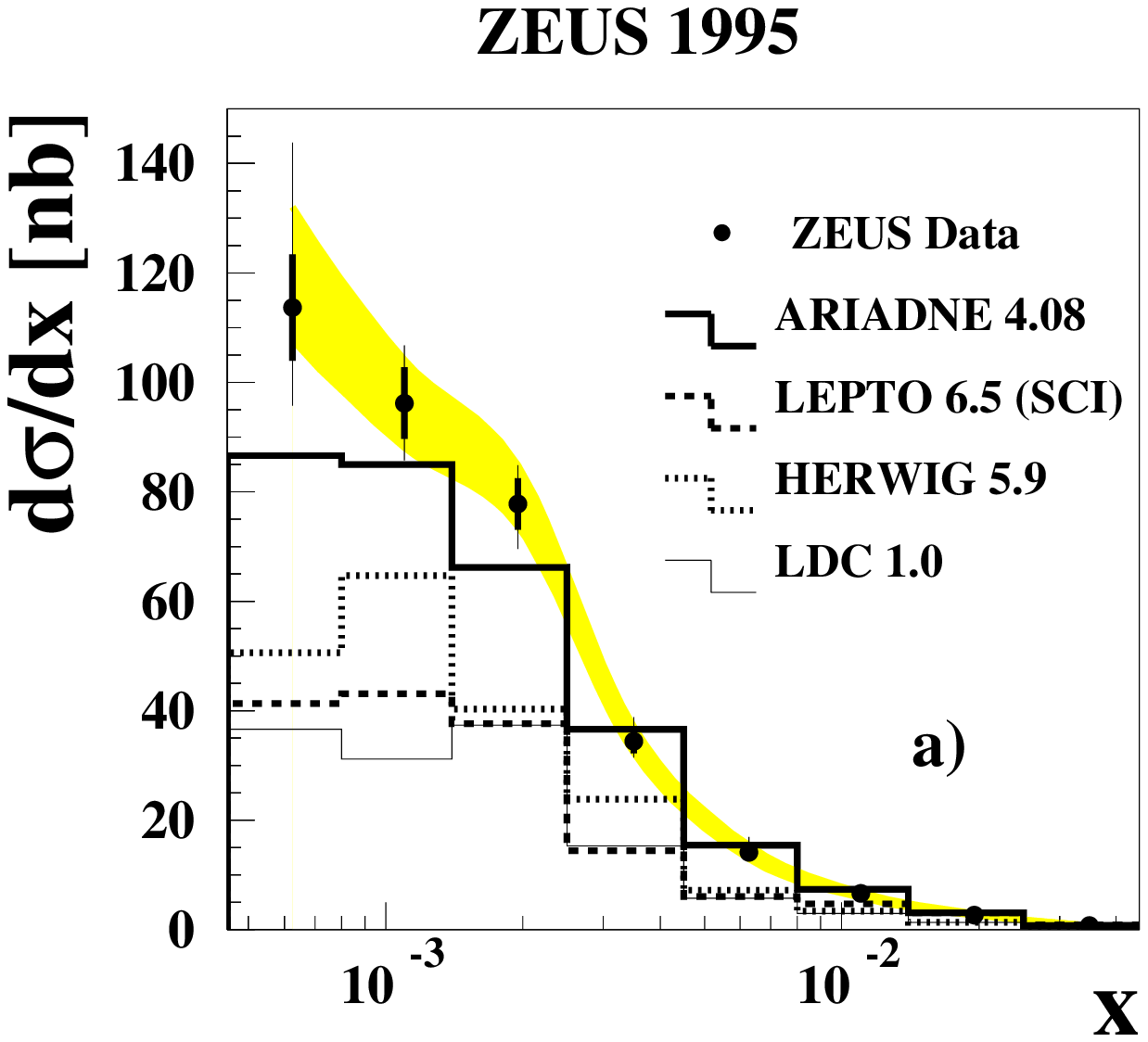,width=6.9cm}} 
  \put(162,-25){\psfig{file=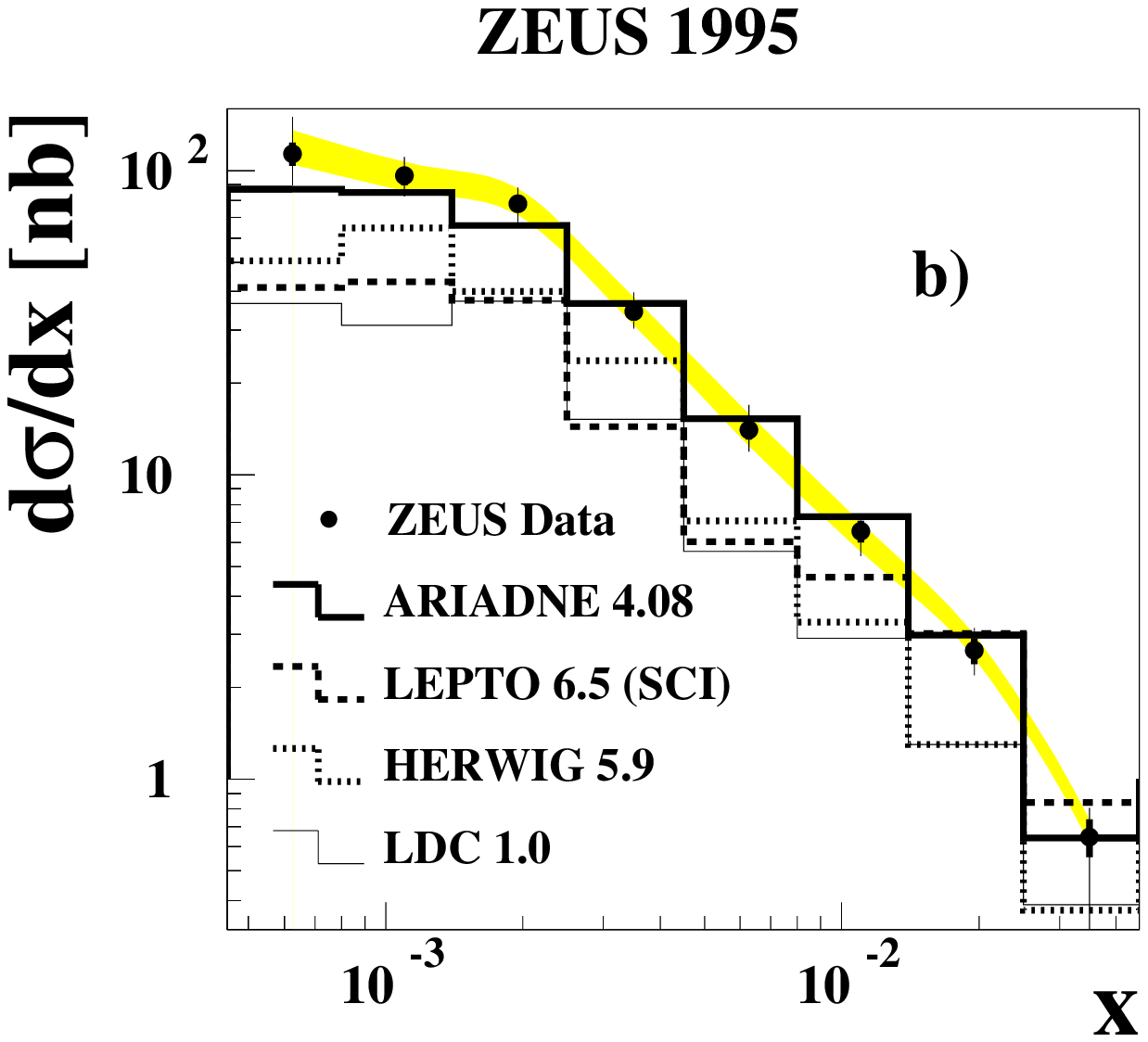,width=6.9cm}} 
 \end{picture}
\caption{Forward jet data of ZEUS \cite{x5} as a function of
  Bjorken-$x$ for $E_{T_{jet}}>5$~GeV. (a) linear scale, (b)
  logarithmic scale. The shaded band gives the error due to the
  uncertainty of the jet energy scale. The data is compared to various
  Monte-Carlo models.\label{ZEUSf}}
\end{figure*}
\begin{figure*}
 \begin{picture}(120,270)
 \put(-5,-50){\epsfig{file=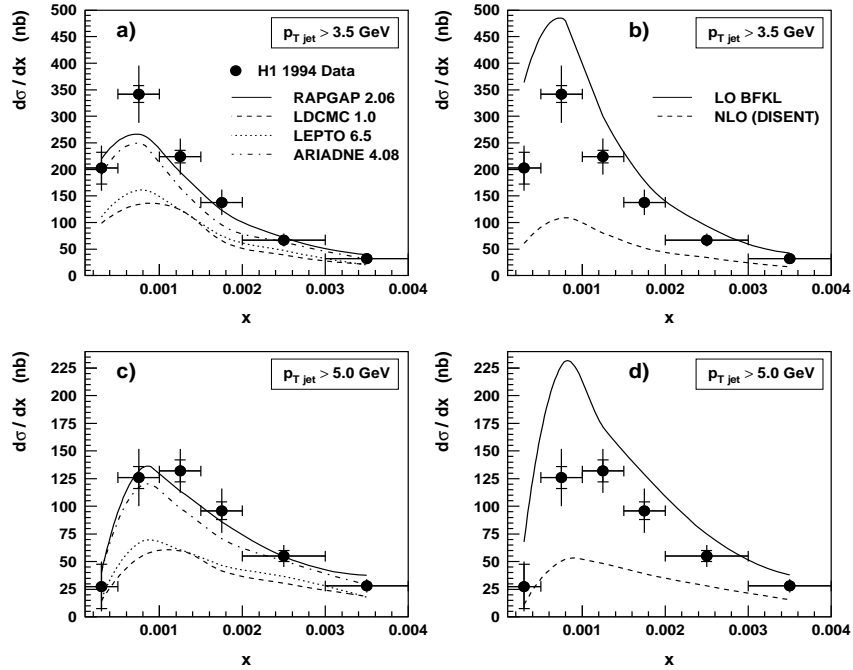,width=14cm,height=28cm,
        bbllx=6pt,bblly=100pt,bburx=647,bbury=1446pt,angle=0,clip=}}
 \end{picture}
\caption{Forward jet data of H1 \cite{x6} as a function of Bjorken-$x$
 for two $E_{T_{jet}}$ cuts of $3.5$~GeV and $5.0$~GeV. (a) and (c)
 contain Monte-Carlo model predictions, whereas (b) and (d) show the 
 results of a LO BFKL (full) and a fixed NLO (dashed) calculation.
 \label{HERAf}} 
\end{figure*}

In Fig.~\ref{ZEUSf} and \ref{HERAf} the collected data from ZEUS and H1 in
the forward region is shown together with predictions from various
Monte-Carlo models. The data from both groups is on the hadron
level. For the ZEUS data, the Monte-Carlo models ARIADNE
\cite{ariadne}, LEPTO \cite{lepto}, HERWIG \cite{herwig} and 
LDC \cite{ldc} have been used for predictions. ARIADNE includes one
of the main features present in the BFKL approach, which is the
absence of the strong $k_T$ ordering. LDC is based on the CCFM
approach and finally LEPTO and HERWIG are based on conventional
leading log DGLAP evolution. Except for HERWIG, the same Monte-Carlo
models are shown in the H1 plots. Instead of HERWIG, the RAPGAP model
\cite{rapgap} is shown, which is also based on DGLAP evolution but
contains an additional resolved virtual photon component. 

As is clear from Fig.~\ref{ZEUSf}, ARIADNE describes the forward jet
cross section reasonably well, apart from the smallest $x$-bin where
it gives slightly too small cross sections. The other three used
models predict cross sections which lie significantly below the
data. Similar results can be extracted from Fig.~\ref{HERAf} (a) and
(c). LDC and LEPTO lie below the data by a factor of 2, whereas
ARIADNE gives a reasonable good description. As an interesting result,
also RAPGAP gives a good description of the H1 data.

\section{BFKL approach}

Of course, attempts have been made to calculate the forward jet cross
section directly within the BFKL formalism. BFKL calculations in
LO by Kwieci\'{n}sky et al., Bartels et al. and Tang \cite{x2} based
on the BFKL approach overshoot the older forward jet data
\cite{x6old}. This can also be seen in Fig.~\ref{HERAf} (b) and (d),
where the LO BFKL calculation on the parton level \cite{x3} (full
line) is compared to the recent H1 data \cite{x6}. These older 
calculations suffer, however, from several deficiencies. They are
asymptotic and do not contain the correct kinematic constraints of the
produced jets. Furthermore they do not allow the implementation of a
jet algorithm as used in the experimental analysis. Also NLO
$\ln(1/x)$ terms in the BFKL kernel \cite{20} predict large negative
corrections which are expected to reduce the forward cross section as
well.

Recently the BFKL calculations have been improved by taking into
account higher order consistency conditions as a way of including
sub-leading corrections to the BFKL equation \cite{x7b}. The
consistency constraint (CC) requires that the virtuality of the emitted
gluons along the chain should arise predominantly from the transverse
components of momentum. By including this CC the authors in \cite{x7b}
claim to recover a dominant part of higher order effects. Furthermore,
the CC is said to subsume energy-momentum conservation over a wide
range of the allowed phase space, which is another source of
sub-leading contributions. 

\begin{figure*}
 \begin{picture}(120,390)
  \put(30,-3){\psfig{file=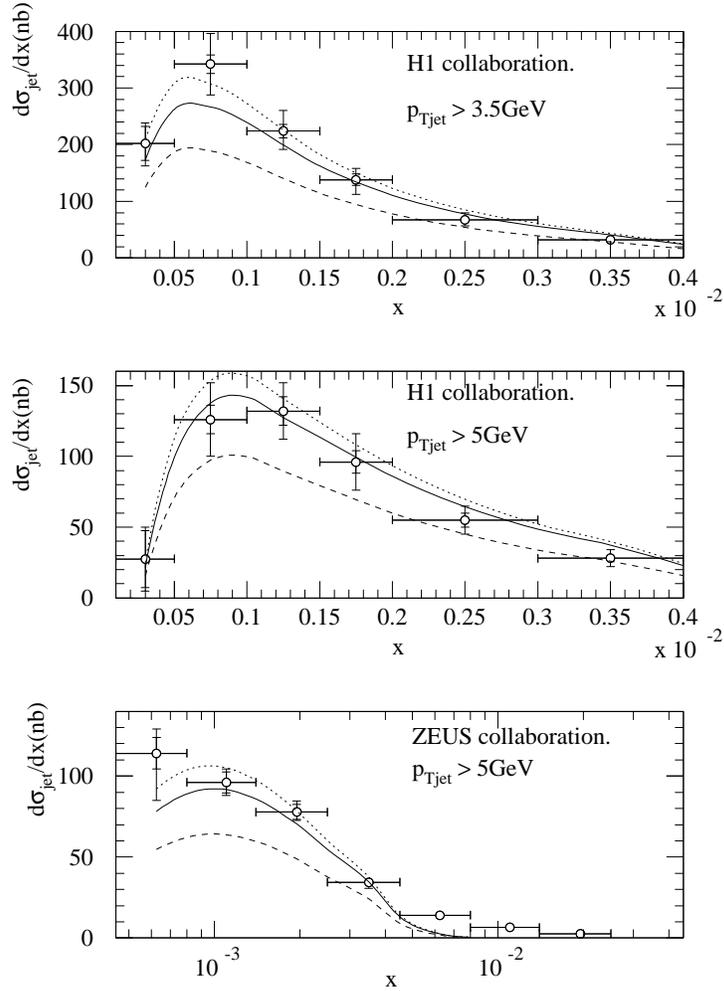,width=10cm}} 
 \end{picture}
\caption{Forward jet data of H1 \cite{x6} and ZEUS \cite{x5} compared
  to prediction based on the BFKL formalism including sub-leading
  corrections \cite{x7b}. The three curves correspond to three
  choices of scales and infrared cut-off. \label{outh}}
\end{figure*}

Including the CC conditions, good agreement between the predictions
and the forward jet data is found, for both the H1 and the ZEUS data,
as shown in Fig.~\ref{outh}. The predictions 
depend on the choice of scale and on an additional infrared cut-off
parameter $k_0$. However, the $k_0$ dependence is much less than the
uncertainty due to the choice of scales. Further details about the
BFKL calculations from \cite{x7b} can be found in these proceedings
\cite{outhi}, also describing the calculation for forward $\pi^0$
production.

\section{Fixed NLO QCD calculations}

Calculations in fixed order perturbation theory have already been
performed for the older H1 forward jet measurement \cite{x6old} by
Mirkes and Zeppenfeld \cite{x7} using their fixed order program MEPJET
\cite{mj}. The calculations where done in next-to-leading order (NLO)
accuracy, i.e., taking the matrix elements up to $O(\alpha_s^2)$ into
account. It was found that the NLO calculations are a factor of 2 to 4
below the data. Similar predictions have been made with help of the
DISENT program \cite{disent} for the more recent H1 measurements
\cite{x6}, shown as the dashed lines in Fig.~\ref{HERAf} (b) and (d),
which confirm the earlier findings. Since the forward jet data can be
succesfully described by the RAPGAP model which includes resolved
virtual photons, it is fair to ask whether also fixed order
calculations including a resolved virtual photon component will be
able to describe the data.

\subsection{Low $Q^2$ jet production in NLO}

NLO calculations for jet production with slightly off-shell direct and
resolved virtual photons have become available recently \cite{kkp},
extending calculations done in the photoproduction regime
\cite{klasen}. The NLO calculations \cite{kkp} are performed with the
phase space slicing method. As is well known the higher order (in
$\alpha_s$) contributions to the direct and resolved cross sections
have infrared and collinear singularities. For the real corrections
singular and non-singular regions of phase space are separated by a
technical cut-off parameter $y_s$. Both, real and virtual corrections,
are regularized by going to $d$ dimensions. The NLO corrections to the
direct process become singular in the limit $Q^2 \rightarrow 0$ in the
initial state on the real photon side. For $Q^2=0$ these photon
initial state singularities are usually also evaluated with the
dimensional regularization method. Then the singular contributions
appear as poles in $\epsilon = (4-d)/2$ multiplied with the splitting
function $P_{q\gamma}$ and have the form
$-\frac{1}{\epsilon}P_{q\gamma}$ multiplied with the LO matrix
elements for quark-parton scattering. These singular contributions are
absorbed into PDF's $f_{a/\gamma }(x)$ of the real photon. For $Q^2
\neq 0$ the corresponding contributions are replaced by  
\begin{equation}
 -\frac{1}{\epsilon} P_{q\gamma} \rightarrow -\ln(s/Q^2) P_{q\gamma}
\end{equation}
where $\sqrt{s}$ is the c.m. energy of the photon-parton subprocess. These
terms are finite as long as $Q^2 \neq 0$ and can be evaluated with $d=4$
dimensions, but become large for small $Q^2$, which suggests
to absorb them as terms proportional to $\ln(M_{\gamma }^2/Q^2)$ in the
PDF of the virtual photon. Parametrizations of the virtual photon have
been provided by several groups \cite{vph}. By this absorption the PDF of
the virtual photon becomes dependent on $M_{\gamma}$, which is the
factorization scale of the virtual photon, in analogy to the real
photon case. Of course, this absorption of large terms is necessary
only for  $Q^2 \ll M_{\gamma}^2$. In all other cases the direct cross
section can be calculated without the subtraction and the additional
resolved contribution. $M_{\gamma}^2$ will be of the order of
$E_T^2$. But also when $Q^2 \simeq M_{\gamma}^2$, one can perform this
subtraction. Then the subtracted term will be added again in the
resolved contribution, so that the sum of the two cross sections
remains unchanged. In this way also the dependence of the cross section
on $M_{\gamma }^2$ must cancel, as long as the resolved contribution 
is calculated in LO only. 

In the general formula for the deep-inelastic scattering cross
section, one has two contributions, the transverse ($d\sigma^U_{\gamma b}$)
and the longitudinal part ($d\sigma^L_{\gamma b}$). Since only the
transverse part has the initial-state collinear singularity the
subtraction in \cite{kkp} has been performed only in the matrix
element which contributes to $d\sigma^U_{\gamma b}$. Therefore  the
longitudinal PDF's $f^L_{a/\gamma }$ are not needed. It is also well known
that $d\sigma^L_{\gamma b}$ vanishes  for $Q^2 \rightarrow 0$. The
calculation of the resolved cross section including NLO corrections
proceeds as for real photoproduction at $Q^2=0$, except
that the cross section is calculated also for final state variables in
the virtual photon-proton center-of-mass system. 

The NLO calculations in the low $Q^2$ region are implemented in
the fixed order program JETVIP \cite{jv}. Various measurements at HERA
in which the jet analysis has been done with JETVIP point to the
presence of a resolved virtual photon component up to moderate
virtualities of $Q^2\simeq 5$~GeV$^2$ \cite{photz}.

\subsection{Comparison to forward jet data}

\begin{figure*}[ttt]
  \unitlength1mm
  \begin{picture}(122,51)
    \put(-3,-50){\epsfig{file=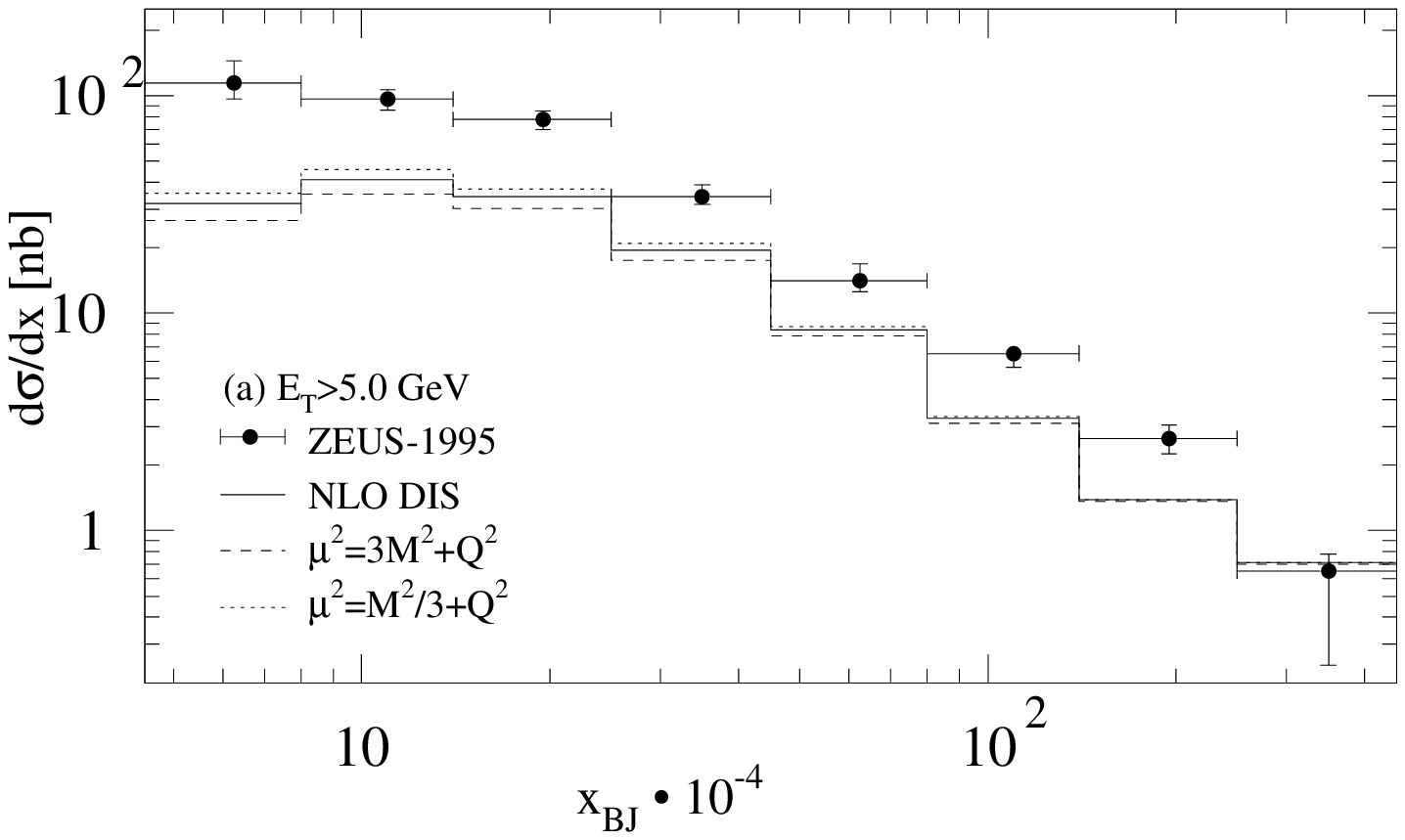,width=6.6cm,height=11cm}}
    \put(58,-50){\epsfig{file=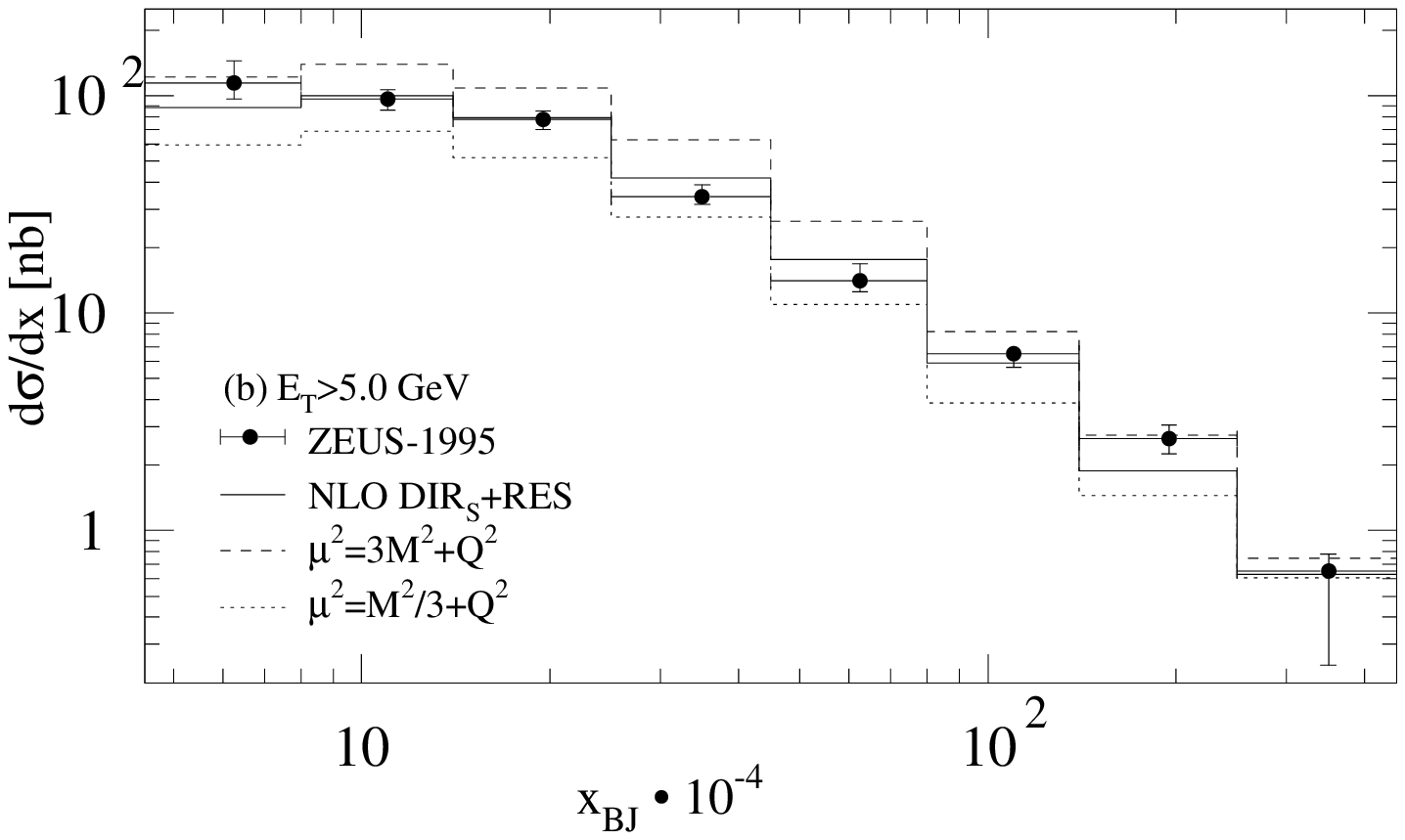,width=6.6cm,height=11cm}}
  \end{picture}
  \begin{picture}(122,51)
    \put(-3,-50){\epsfig{file=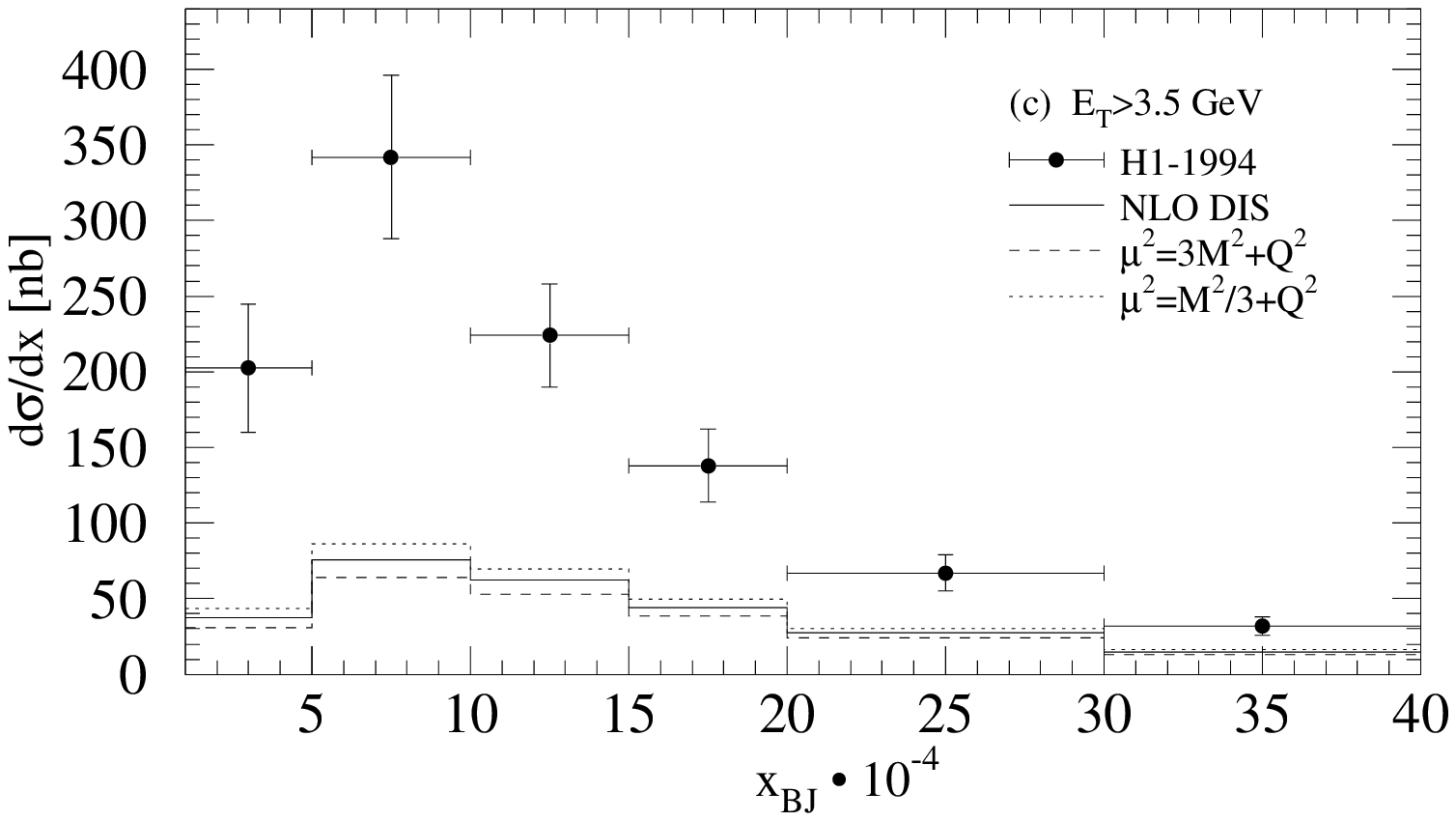,width=6.6cm,height=11cm}}
    \put(58,-50){\epsfig{file=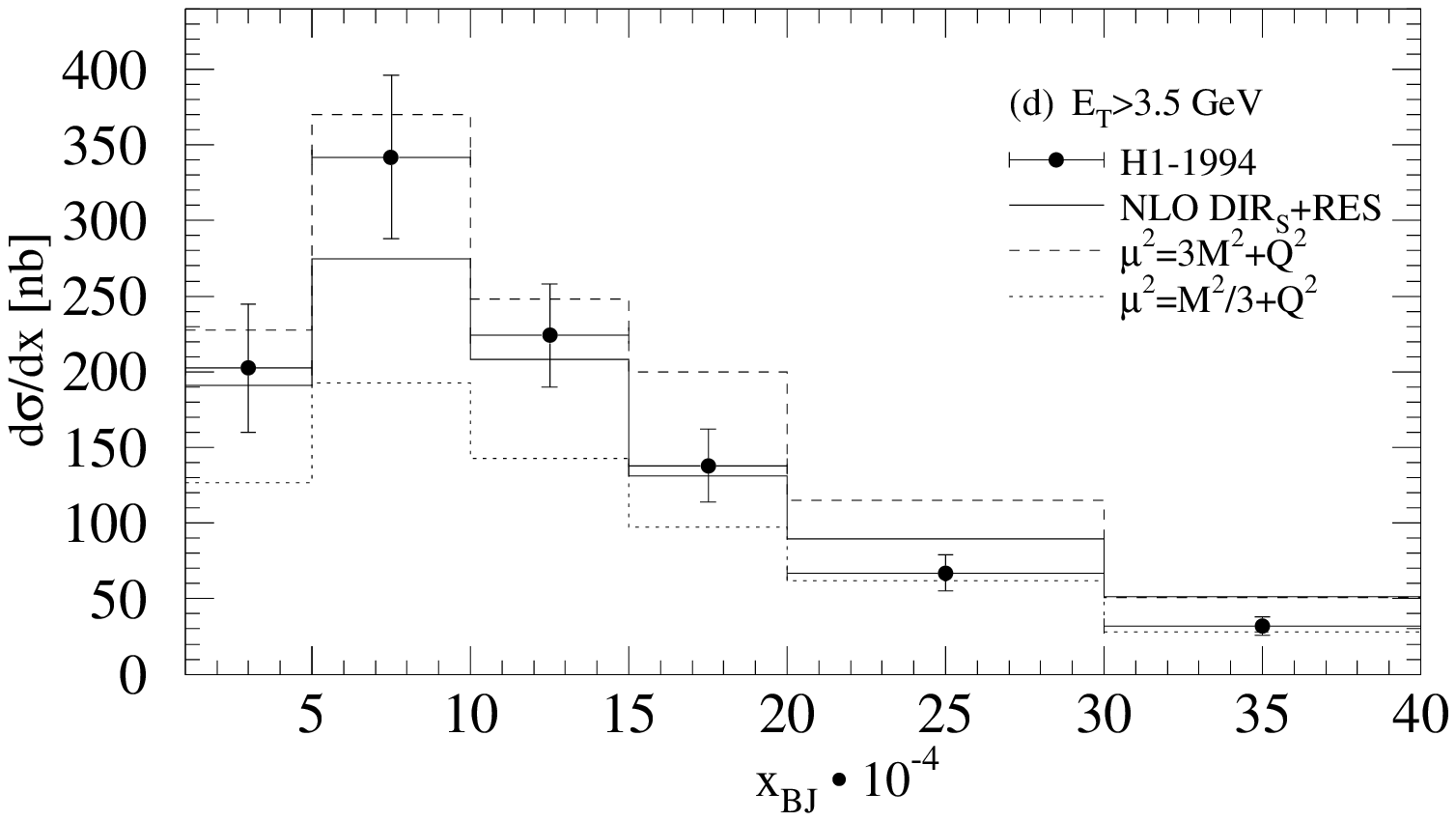,width=6.6cm,height=11cm}}
  \end{picture}
\caption{Dijet cross section in the forward region compared to HERA
data: (a) and (b) ZEUS; (c) and (d) H1.
(a) NLO DIS, $E_T>5$ GeV; (b) NLO DIR$_S$+RES, $E_T>5$ GeV;
(c) NLO DIS, $E_T>3.5$ GeV; (d) NLO DIR$_S$+RES, $E_T>3.5$ GeV.
\label{xkp1}}
\end{figure*}

Recently, we have performed a NLO calculation including the virtual
resolved photon for the forward jet region \cite{xkp} with the help of
JETVIP. The results for the ZEUS kinematical conditions are shown in
Fig.~\ref{xkp1}~a,b. In Fig.~\ref{xkp1}~a we plotted the full ${\cal
O}(\alpha_s^2)$ inclusive two-jet cross section (DIS) as a function
of $x$ for three different scales $\mu^2=\mu_R^2=3M^2+Q^2, M^2+Q^2$
and $M^2/3+Q^2$ with a fixed $M^2=50$~GeV$^2$ related to the mean
$E_T^2$ of the forward jet and compared them with the measured points
from ZEUS \cite{x5}. The choice $\mu_F^2 > Q^2$ is mandatory if we
want to include a resolved contribution. Similar to the results
obtained with MEPJET and DISENT, the NLO direct cross section is by a
factor 2 to 4 too small compared to the data. The variation inside the
assumed range of scales is small, so that also with a reasonable
change of scales we can not get agreement with the data. In
Fig.~\ref{xkp1}~b we show the corresponding forward jet cross sections
with the NLO resolved contribution included, labeled DIR$_S$+RES,
again for the  three different scales $\mu $ as in
Fig.~\ref{xkp1}~a. Now we find good agreement with  the ZEUS data. The
scale dependence is not so large that we must fear our results not to
be trustworthy.

In Fig.~\ref{xkp1}~c,d we show the results compared to the H1 data \cite{x6}
obtained with $E_T>3.5$~GeV in the HERA system. In the plot on the left 
the data are compared with the pure NLO direct prediction, which turns
out to be too small by a similar factor as observed in the comparison
with the ZEUS data. In Fig.~\ref{xkp1}~d the forward jet cross section is
plotted with the NLO resolved contribution included in the way
described above. We find good agreement with the H1 data inside
the scale variation window $M^2/3+Q^2<\mu^2 < 3M^2+Q^2$. We have also
compared the predictions with the data from the larger $E_T$ cut, namely
$E_T>5.0$~GeV, and found similar good agreement \cite{xkp}.

As described in \cite{xkp} the NLO resolved contribution supplies higher
order terms in two ways, first through the NLO corrections in the hard
scattering cross section and second in the leading logarithmic
approximation by evolving the PDF's of the virtual photon to the
chosen factorization scale. This way the logarithms in
$E_T^2/Q^2$ are summed, which, however, in the considered kinematical 
region is not an important effect numerically. Therefore, the
enhancement of the NLO direct cross section through inclusion of
resolved processes in NLO is mainly due to the convolution of the
point-like term in the photon PDF with the NLO resolved matrix
elements, which gives an approximation to the NNLO direct cross
section without resolved contributions. One can therefore speculate
that the  forward jet cross section could be described within a fixed
NNLO calculation, using only direct photons.

\section{Conclusions}

We conclude that two alternative models exist for describing the
forward jet data. The BFKL calculation including CC describes the data
well. This is supportet by the ARIADNE model, where strong $k_T$
ordering is absent and which also describes the data. It is however
not clear, why the LDC model does not decribe the data, although BFKL
dynamics should be included in the respective regime of validity. 

Similarly, the NLO theory with a resolved virtual photon
contribution as an approximation of the NNLO DIS cross section, which
is presently not available, gives a good description of the forward
jet data. Assumably, the RAPGAP model produces these higher order
effects through parton shower contributions in the resolved cross 
section.

\subsubsection*{Acknowledgments.}

I thank the organizers for the kind invitation and the pleasant workshop
athmosphere. The results of section 4 where obtained in collaboration
with G.~Kramer.


\clearpage
\addcontentsline{toc}{section}{Index}
\flushbottom
\printindex

\begin{thebibliography}{99}
\addcontentsline{toc}{section}{References}
\bibitem{sxH1} H1 Collaboration (I Abt et al.)
\Journal{\NPB}{407}{515}{1993}; H1 Collaboration (T Ahmed et al.)
 \Journal{\NPB}{439}{471}{1996}

\bibitem{sxZeus} Zeus Collaboration (M Derrik et al.)
\Journal{\PLB}{316}{412}{1993}; \Journal{\ZPC}{65}{379}{1995}; 
\Journal{\ZPC}{69}{607}{1996}

\bibitem{dglap} V N Gribov, L N Lipatov, Sov. J. Nucl. Phys. {\bf 15}
(1972) 438, 675; Y L Dokshitzer, Sov. Phys. JETP {\bf 46} (1977)
641;  G Altarelli, G Parisi, Nucl. Phys. {\bf B126} (1977) 298

\bibitem{BFKL}
E A Kuraev, L N Lipatov, Y S Fadin, Sov. Phys. JETP {\bf 45} (1977) 199;
Ya Ya Balitzky, L N Lipatov, Sov. J. Nucl. Phys. {\bf 28} (1978) 288 

\bibitem{ccfm} M Ciafaloni, \Journal{\NPB}{296}{49}{1988}; S Catani,
F Fiorani, G Marchesini, \Journal{\PLB}{234}{339}{1990}, 
\Journal{\NPB}{336}{18}{1990}; G Marchesini, \Journal{\NPB}{445}{49}{1995} 

\bibitem{x1} A Mueller, Nucl. Phys. B (Proc. Suppl.) {\bf 18C} (1990) 125;  
J. Phys. {\bf G17} (1991) 1443

\bibitem{x5}
ZEUS Collaboration (J Breitweg et al.) \Journal{\EPJC}{6}{239}{1999}

\bibitem{x6}
H1 Collaboration (C Adloff et al.) \Journal{\NPB}{538}{3}{1999}

\bibitem{ariadne} L L\"onnblad, Comp. Phys. Comm. {\bf 71} (1992) 15

\bibitem{lepto} G Ingelman, A Edin, and J Rathsman,
Comp. Phys. Comm. {\bf 101} (1997) 108 

\bibitem{herwig} G Marchesini et al., Comp. Phys. Comm. {\bf 82}
(1992) 445

\bibitem{ldc} H Kharraziha, L L\"onnblad, JHEP {\bf 9803} (1998) 6

\bibitem{rapgap} H Jung, Comp. Phys. Comm. {\bf 86} (1995) 147;\\
H Jung, L J\"onsson, H K\"uster, \Journal{\EPJC}{9}{383}{1999}

\bibitem{x2}
J Kwieci\'{n}sky, A D Martin, J P Sutton, \Journal{\PRD}{46}{921}{1992};\\
J Bartels, A De Roeck, M Loewe, \Journal{\ZPC}{54}{635}{1992};\\
W K Tang, \Journal{\PLB}{278}{363}{1992}

\bibitem{x6old}
H1 Collaboration (S Aid et al.) \Journal{\PLB}{356}{118}{1995}

\bibitem{x3}
J Bartels, V Del Duca, A De Roeck, D Graudenz, M W\"usthoff,
\Journal{\PLB}{384}{300}{1996}

\bibitem{20} V S Fadin, L N Lipatov, \Journal{\PLB}{429}{127}{1998}

\bibitem{x7b} J Kwieci\'{n}sky, A D Martin, J J Outhwaite,
\Journal{\EPJC}{9}{611}{1999} 

\bibitem{outhi} J J Outhwaite, these proceedings 

\bibitem{x7}
E Mirkes, D Zeppenfeld, \Journal{\PRL}{78}{428}{1997}

\bibitem{mj}
E Mirkes, D Zeppenfeld, \Journal{\PLB}{380}{205}{1996}

\bibitem{disent} S Catani, M H Seymour, \Journal{\PLB}{378}{287}{1996};
\Journal{\NPB}{485}{291}{1997}

\bibitem{kkp} M Klasen, G Kramer, B P\"otter, Eur. Phys. J. {\bf C1}
(1998) 261; \\ G Kramer, B P\"otter,
Eur. Phys. J. {\bf C5} (1998) 665; \\
B P\"otter, Eur. Phys. J. Direct {\bf C5} (1999) 1, hep-ph/9707319  

\bibitem{klasen} M Klasen, these proceedings, hep-ph/9907366

\bibitem{vph} M Gl\"uck, E Reya, M Stratmann,
\Journal{\PRD}{51}{3220}{1995}; G A Schuler, T Sj\"ostrand,
\Journal{\ZPC}{68}{607}{1995}; \Journal{\PLB}{376}{193}{1996};
M Gl\"uck, E Reya, I Schienbein, \Journal{\PRD}{60}{054019}{1999}

\bibitem{jv} B P\"otter, Comp.~Phys.~Comm. {\bf 119} (1999) 45

\bibitem{photz} S Maxfield, B P\"otter, L Sinclair, J. Phys. {\bf G25}
(1999) 1465; \\ G Kramer, B P\"otter, Lund-Workshop, Sweden 1998,
ed. G~Jarlskog and T~Sj\"ostrand, p.~29, hep-ph/9810450;
B P\"otter, DIS98 Workshop, Belgium 1998, ed. Gh~Coremans and
R~Roosen, p.~574, hep-ph/9804373 

\bibitem{xkp} G Kramer, B P\"otter, Phys. Lett. {\bf B453} (1999) 295

\end{thebibliography}
\end{document}